\documentclass[twocolumn,pra, noshowpacs, noshowkeys,nofootinbib]{revtex4}

\usepackage{latexsym}
\usepackage{amssymb}
\usepackage{graphicx}
\usepackage{amsmath}
\usepackage{amsfonts}
\usepackage[english]{babel}
\usepackage{epsfig}
\usepackage{subfigure}
\usepackage[T1]{fontenc}
\usepackage{graphics}
\usepackage{graphicx, lscape}
\usepackage{supertabular}
\usepackage{eucal,amsfonts}
\usepackage{amsbsy}

\usepackage{color}

\def\espace{\hspace{2mm}}
\def\Tr{{\rm Tr}\,}
\def\rank{{\rm rank}\,}

\def\k(#1){|#1\rangle}
\def\b(#1){\langle#1|}
\def\bk(#1,#2){ \langle #1 | #2 \rangle}
\def\C(#1){{\cal#1}}

\graphicspath{{./Figure/}}

\begin{document}

\title{Efficient Optimal Minimum Error Discrimination of\\
Symmetric Quantum States}
\date{\today}

\author{Antonio \surname{Assalini}}
\author{Gianfranco \surname{Cariolaro}}
\author{Gianfranco \surname{Pierobon}}

\affiliation{Department of Information Engineering (DEI), University
of Padua, \\ Via Gradenigo 6/B, 35131, Padova, Italy}

\date{January, 2010}

\begin{abstract}
This paper deals with the quantum optimal discrimination among mixed
quantum states enjoying geometrical uniform symmetry with respect to
a reference density operator $\rho_0$. It is well-known that the
minimal error probability is given by the positive operator-valued
measure (POVM) obtained as a solution of a convex optimization
problem, namely a set of operators satisfying geometrical symmetry,
with respect to a reference operator $\Pi_0$, and maximizing
$\textrm{Tr}(\rho_0 \Pi_0)$. In this paper, by resolving the dual
problem, we show that the same result is obtained by minimizing the
trace of a semidefinite positive operator $X$ commuting with the
symmetry operator and such that $X \geq \rho_0$. The new formulation
gives a deeper insight into the optimization problem and allows to
obtain closed-form analytical solutions, as shown by a simple but
not trivial explanatory example. Besides the theoretical interest,
the result leads to semidefinite programming solutions of reduced
complexity, allowing to extend the numerical performance evaluation
to quantum communication systems modeled in Hilbert spaces of large
dimension.
\end{abstract}

\pacs{03.65.Ta, 03.67.Hk,  05.30.Ch}

\keywords{Should be Optional }


 \maketitle


\section{Introduction}\label{Sect:1}

 In a quantum system Alice prepares the quantum channel
into one of several quantum states. Bob measures the quantum channel
by a set of measurement operators and, on the basis of the result,
it guesses the choice made by the transmitter. These actions lead to
a classical channel, and the problem arises of finding the
measurement operators that provide optimal performance according to
a predefined criterion, in this paper the minimum error probability.
However, to solve the problem, except for some particular cases, has
appeared to be a very difficult task since the pioneering
contributions in the seventies~\cite{Helstrom, Holevo, Yuen}.

In recent years, particular attention has been paid to quantum
states satisfying geometrical symmetry \cite{Ban, Eldar1, Eldar2,
Kato}, in view of applications to optical communication systems. In
some specific cases, including symmetric pure quantum states
\cite{Ban, Eldar1} and symmetric mixed quantum states with a
characteristic structure \cite{Eldar2}, the solution named square
root measurement (SRM) proves optimal. Nevertheless, in general, SRM
represents a suboptimal strategy, although it provides pretty good
performance in many scenarios.

In this paper, we are concerned with the construction of optimal
POVM for the discrimination of symmetric \textit{mixed} quantum
states. We present some results that provide intuition into the
problem and offer perspectives on its solution.

Optimal quantum state discrimination represents a convex
optimization problem, and, as such, it can be formulated in a
\textit{primal} and in a \textit{dual} form, with the latter having
a reduced number of variables and constraints \cite{Yuen, Eldar2}.
In this paper we investigate how primal and dual problems simplify
with symmetric quantum states. For the primal problem this study was
already considered in \cite{Eldar2}. Herein, we extend the analysis
to the dual problem, where the optimal solution can be searched in a
set of smaller dimension. The simplified formulation of the dual
problem is illustrated with an example where a closed-form solution
is easily found. For problems of large dimension, that cannot be
solved analytically, the reduced number of variables in the
simplified dual statement becomes useful to the numerical solution
by means of semidefinite programming (SDP) tools.


\section{General Formulation of Minimum error discrimination} \label{Sect:2}

The quantum decision problem is formalized in a $N$-dimensional
complex Hilbert space $\mathcal{H}$ \cite{Helstrom}, where an
ensemble of quantum states $\rho_i$, $i=0,1,\ldots,M-1$, with prior
probabilities $q_i \geq 0$, $\sum_{i=0}^{M-1} q_i=1\,$, is given.
The quantum states $\rho_i$ are density operators on $\C(H)$, i.e.,
(self-adjoint) positive semidefinite (PSD) operators ($\rho_i\ge0$),
with unit trace, ${\Tr(\rho_i)=1}$. For notation convenience, we
denote by $\mathcal{P}$ the class of the PSD operators on
$\mathcal{H}$.
The measurement operators $\Pi_i$, $i=0,1,\ldots,M-1$, constitute a
POVM having the properties $\Pi_i\in \mathcal{P}$ and
$\sum_{i=0}^{M-1}\Pi_i=I\,,$ where $I$ is the identity operator on
$\C(H)$. The transition probabilities of the resulting quantum
channel become ${p\,(j|i)=\Tr(\rho_i\Pi_j)}$, so that the
probability of correct detection is given by $
P_c=\sum_{i=0}^{M-1}q_i\Tr(\rho_i\Pi_i)\,$.

Hence, the problem of finding the maximum probability of correct
state discrimination can be concisely stated as follows.\\[-2mm]

\noindent\textbf{Primal problem (\texttt{PP1}).} Find the maximum of
the probability of correct detection
$P_c=\sum_{i=0}^{M-1}q_i\Tr(\rho_i\Pi_i)$ over the class of the
POVM on $\mathcal{H}$.\\[-2mm]

The analytical solution of \texttt{PP1} is in general difficult
since $P_c$ has to be maximized over the whole $M$-tuple of measure
operators $\Pi_i$. As a matter of fact, closed-form results to the
primal problem are available only for some particular quantum
mechanical systems, e.g., the binary case \cite{Helstrom}.
Nevertheless, since the objective is to search a global maximum of a
linear function into a convex set, the problem can be faced by means
of numerical tools such as SDP. Besides, according to classical
results in convex optimization theory \cite{Boyd}, in place of the
primal problem, it is in general more convenient to consider its
corresponding dual problem, since it presents
a smaller number of variables and constraints \cite{Yuen,Eldar2}.\\[-2mm]

\noindent\textbf{Dual problem (\texttt{DP1}).} Minimize the trace of
the optimization operator $X$ over the class $\mathcal{P}$, subject
to the constraints $X\geq q_i \rho_i$, $i=0,1,\ldots,M-1$. Once
found a minimum trace operator $X_{\textrm{opt}}$, its trace gives
the maximum probability of correct detection, $P_c=\Tr(X_{\textrm{opt}})$.\\[-2mm]

The equalities
$(X_{\textrm{opt}}-q_i\rho_i)\Pi_i=\Pi_i(X_{\textrm{opt}}-q_i\rho_i)=0$,
$i=0,1,\ldots,M-1$, are necessary conditions on the optimal POVM.
These conditions become sufficient, once the searched measure
operators are constrained to belong to $\mathcal{P}$ and to solve
the identity on $\mathcal{H}$.


\section{Discrimination of Symmetric Quantum States}\label{Sect:3}

In quantum detection an important role is played by
\textit{geometrically uniform symmetry} \cite{Ban,Eldar1,Eldar2}.
Among the several generalizations, we consider the basic case of
symmetric mixed quantum states generated from a reference density
operator $\rho_0$ as
\begin{equation}
\rho_i = S^{i}\, \rho_0\, S^{-i} \quad,\quad i=0,1,\ldots,M-1\quad,
\label{eq:5}
\end{equation}
where the \textit{symmetry operator} $S$ is unitary
({$SS^\dag=S^\dag S=I$}) and such that $S^M=I$. The geometry
implicitly requires that the mixed states are equiprobable, i.e.,
$q_i=1/M$, $i=0,1,\ldots,M-1$. In  \cite{Eldar2} it was shown that
optimal POVM having the same symmetry can always be found. Hence,
without loss of generality, we can assume that
\begin{equation}
\Pi_i = S^{i}\, \Pi_0\, S^{-i}   \quad,\quad  i=0,1,\ldots,M-1\quad,
\label{eq:6}
\end{equation}
where $\Pi_0$ is the reference measure operator. Consequently, for a
fixed $M$, the knowledge of $\rho_0$, $\Pi_0$ and $S$ is sufficient
to fully describe the state ensemble and the POVM.

The density operators \eqref{eq:5} have all the same rank, and the
same holds for the measure operators \eqref{eq:6}. As proved in
\cite{Eldar3}, the optimal measure operators can be assumed to have
rank no higher than that of the corresponding density operators,
namely $\rank(\Pi_0)\leq\rank(\rho_0)\,$.


\subsection{Primal Problem for Symmetric Quantum States}\label{Sect:3a}

The specific geometry of the state ensemble can be exploited to get
insight into how solving the state discrimination problem. In
particular, the primal problem \texttt{PP1} can be rewritten in a
simpler form as follows.

\noindent\textbf{Primal problem for symmetric quantum states
(\texttt{PP2}).} Find the maximum of $P_c=\Tr(\rho_0\,\Pi_0)$, with
$\Pi_0\in\mathcal{P}$ and such that
$\sum_{i=0}^{M-1} S^i \, \Pi_0 \,S^{-i}= I$.\\
\textbf{Proof:} This formulation was first given in \cite{Eldar2}.
It can straightforwardly be proved by using \eqref{eq:5} and
\eqref{eq:6} in \texttt{PP1}, so that $P_c=\sum_{i=0}^{M-1} q_i
\Tr(\rho_i\,\Pi_i)=\sum_{i=0}^{M-1} \frac{1}{M}  \Tr(S^i \rho_0
S^{-i}\,S^i \Pi_0 S^{-i})=\Tr( \rho_0 \,\Pi_0)$. Moreover, if $\Pi_0
\geq 0$ then $\Pi_i =S^i \Pi_0 S^{-i}\geq 0\,$. $\blacksquare$


\subsection{Dual Problem for Symmetric Quantum States}\label{Sect:3b}

The optimization problem \texttt{PP2} is comparatively simple, and,
perhaps, this is the reason why no particular attention has been
paid in the literature to the study of the dual theorem to obtain an
alternative formulation. In the following we investigate this point.\\[-2mm]

\noindent\textbf{Dual problem for symmetric quantum states
(\texttt{DP2}).} Minimize the trace of the optimization operator $X$
over the class $\mathcal{P}$, subject to the constraints $X \geq
\frac{1}{M}\rho_0$ and $XS=SX$. Once found a minimum trace operator
$X_{\textrm{opt}}$, its trace gives
the maximum probability of correct detection, $P_c=\Tr(X_{\textrm{opt}})$.\\
\textbf{Proof:} Define $\rho_i'=S^i \big(\frac{1}{M} \rho_0\Big)
S^{-i}$. Let $\mathcal{V}$ be the feasible set according to the
general dual problem in Section~\ref{Sect:2}, i.e., the set of PSD
operators $X$ such that $X \geq \rho_i'$, $i=0,1,\ldots,M-1$, and
let $\mathcal{V}'$ be the set of PSD operators $X'$ such that $X'
\geq \rho_0'$ and $X'S=SX'$. The proof is organized in two steps. In
the first step it is shown that $\mathcal{V}' \subset \mathcal{V}$,
while in the second step it is proved that for any $X \in
\mathcal{V}$ there exist $X' \in \mathcal{V}'$ such that
$\Tr(X')=\Tr(X)$. Then, the
search of $X$ can be confined into $\mathcal{V}'$.\\
{- \textit{Step 1}:} If $X' \in \mathcal{V}'$, for the commutativity
between $X'$ and $S$ we get $X'=S X' S^{-1}$ and recursively $X'=S^i
X' S^{-i}$. Hence, for any $i$
\begin{equation}
\begin{split}
X'-\rho_i'&=X'- S^i \rho_0' S^{-i}\\&=S^i X' S^{-i} - S^i \rho_0'
S^{-i}\\& =S^i (X' - \rho_0') S^{-i} \geq 0\;,
\end{split}
\end{equation}
since $X' \geq \rho_0'$ for assumption.\\
{- \textit{Step 2}:} For each $X \in \mathcal{V}$ we consider
$$
X' = \frac{1}{M} \sum_{i=0}^{M-1} S^{-i}X S^i\;.
$$
Being $X\geq \rho_i'$ for each $i$, it follows that
$$
X' \geq \frac{1}{M} \sum_{i=0}^{M-1} S^{-i} \rho_i' S^{i}= \rho_0'
\:.
$$
Moreover, recalling that $S^M=I$
$$
SX'S^{-1}= \frac{1}{M}\sum_{i=0}^{M-1} S^{-(i-1)} X S^{i-1}= X'
$$
and then $X'$ commutes with $S$. Finally,
\begin{equation}
\begin{split}
 \Tr(X')&=  \frac{1}{M} \sum_{i=0}^{M-1} \Tr(S^{-i}X S^i)
=\sum_{i=0}^{M-1} \Tr(X S^i S^{-i})\\&= \frac{1}{M} \sum_{i=0}^{M-1}
\Tr(X)= \Tr(X) \;,\\[-5mm]
\end{split}\nonumber
\end{equation}
and the proof is complete. $\blacksquare$\\[-2mm]

Therefore, the search of the unknown optimization operator $X$ can
be restricted to the subclass of $\mathcal{P}$ composed by the PSD
operators that commute with the symmetry operator~$S$.

The optimal $\Pi_0$ can be found by the relations
$(X_{\textrm{opt}}-\frac{1}{M}\rho_0)\Pi_0=\Pi_0(X_{\textrm{opt}}-\frac{1}{M}\rho_0)=
0$ subject to $\Pi_0 \in \mathcal{P}$ and $\sum_{i=0}^{M-1} S^i \,
\Pi_0 \,S^{-i}= {I}\,$.

In the next section, we develop a formulation, where the commutation
condition $XS=SX$ is replaced by an alternative
constraint.\vspace{-1mm}
\subsection{Alternative Formulation of the Dual Problem}\label{Sect:3c}\vspace{-1mm}
Since the symmetry operator $S$ is known a-priori, it is possible to
exploit its spectral characterization to rewrite the dual problem
\texttt{DP2} as follows.\\[-2mm]

\noindent\textbf{Other form for the dual problem for symmetric
quantum states (\texttt{DP3}).} Let
$\lambda_0,\lambda_1,\ldots,\lambda_{\bar{N}-1}$ be the $\bar{N}\leq
N$ \textit{distinct} eigenvalues of $S$, and $N_i$ be the
multiplicity of $\lambda_i$. Let $U$ be a basis of eigenvectors of
$S$, with the first $N_0$ eigenvectors corresponding to $\lambda_0$,
the next $N_1$ eigenvectors corresponding to $\lambda_1$, and so on.
Then, minimize the trace of the optimization operator $\tilde{X}$
over the subclass of $\mathcal{P}$ consisting of
\textit{block-diagonal} operators with blocks of dimension $N_i$,
under the constraint $\tilde{X} \geq \frac{1}{M} U^\dag \rho_0 U$.
Once found a minimum trace operator $\tilde{X}_{\textrm{opt}}$, its
trace gives the maximum probability
of correct detection, $P_c=\Tr(\tilde{X}_{\textrm{opt}})$.\\
\textbf{Proof:} By \texttt{DP2} the optimization operator $X$ can be
chosen to commute with $S$. Therefore, given an eigenbasis $U$ for
$S$, we can write $X=U \tilde{X} U^\dag$ where, by well-known
results on simultaneous diagonalization of commuting self-adjoint
operators \cite{Sadun}, $\tilde{X}$ turns out to be block diagonal
with the size of the blocks given by the multiplicity of the
eigenvalues of $S$. Moreover, we find that ${\Tr(X)=\Tr(U \tilde{X}
U^\dag)=\Tr(\tilde{X})}$ and the constraint $X \geq
\frac{1}{M}\rho_0$ becomes $\tilde{X} \geq \frac{1}{M} U^\dag \rho_0
U$. $\blacksquare$\\[-2mm]

The commutative requirement $XS=SX$ is then replaced by fixing the
block-diagonal structure on $\tilde{X}\,$. Given an optimal
$\tilde{X}_{\textrm{opt}}$, the reference optimal operator $\Pi_0$
is solution of $(U \tilde{X}_{\textrm{opt}}
U^\dag-\frac{1}{M}\rho_0)\Pi_0=\Pi_0 (U \tilde{X}_{\textrm{opt}}
U^\dag-\frac{1}{M}\rho_0) = 0$, subject to $\Pi_0\in \mathcal{P}$
and $\sum_{i=0}^{M-1} S^i \, \Pi_0 \,S^{-i}= {I}\,$.

The new formulation of the dual problem is given in a form that is
particularly suitable for SDP computational tools \cite{Grant} and,
moreover, it is analytically more tractable that the previous
version. We now  quantify the complexity of the different
approaches.


\begin{center}
\begin{table}[t]
\begin{minipage}{8.5cm}
\caption{\label{Table:1} Number of real decision variables $d$,
equality constraints~$C_e$ and inequality constraints $C_i$ for the
optimization problems.\\[-2mm]}
\begin{minipage}{7cm}
\begin{ruledtabular}
\begin{tabular}{c||ccc}
\hspace{12mm}    &      \hspace{3mm}   $d$ \hspace{3mm} &
\hspace{3mm}$C_e$\hspace{3mm}    & \hspace{3mm}$C_i$\hspace{5mm}
\\ \hline
\texttt{PP1} &        $MN^2$                   &    1       &     $M$      \\
\texttt{PP2} &        $N^2$                    &    1       &      1       \\
\texttt{DP1} &        $N^2$                    &    0       &     $M$      \\
\texttt{DP2} &        $N^2$                    &    0       &      2       \\
\texttt{DP3} &   $\sum_{i=0}^{\bar{N}} N_i^2$  &    0       &      1       \\
\end{tabular}
\end{ruledtabular}
\end{minipage}
\end{minipage}\vspace{-10mm}
\end{table}
\end{center}

\subsection{Problem dimension and number of constraints}\label{Sect:3d}
The complexity of a linear program, to get the solution of a convex
optimization problem, is hard to evaluate in terms of arithmetic
operations (see \cite{Boyd} for details). Nevertheless, since we are
seeking a global optimum, the dimension of the feasible region for
the objective function gives an order of the complexity of the
problem \cite{Boyd}.

The set of self-adjoint operators on $\mathcal{H}$ forms an
$N^2$--dimensional real vector space. Therefore, for a given
optimization problem we can find the dimension $d$ of the
correspondent real space on which the considered objective function
is defined. In other terms, we find the number of real variables $d$
in a given objective function. The results are summarized in Table
\ref{Table:1} where $C_e$ and $C_i$ represent, respectively, the
number of equality and inequality constraints for a given problem.
Note that the PSD condition on self-adjoint operators is counted as
an inequality, e.g., the relation $\Pi_0 \in \mathcal{P}$ is counted
as the inequality $\Pi_0\geq 0$.

The dual problem \texttt{DP3} presents the smaller number of
variables and constraints among the considered optimization problems
and, in particular, $d=\sum_{i=0}^{\bar{N}-1} N_i^2$ is smaller than
$N^2$,  depending on the spectrum of the symmetry operator $S$. If
$S$ has all distinct eigenvalues, then $\bar{N}=N$, $N_i=1$ for each
$i$, and the optimization operator $\tilde{X}$ in \texttt{DP3}
becomes diagonal, giving $d=N< N^2\,$.


\section{Example of Application}\label{Sect:4}
In this section, the previous results are applied to the
discrimination of an ensemble of $M$ symmetric mixed quantum states
on a 2-dimensional ($N=2$) complex Hilbert space. The symmetry
operator is
\begin{equation}
S=\left[
\begin{array}{ccc}
\cos\left(\frac{\pi}{M}\right) &\;&
-\sin\left(\frac{\pi}{M}\right)\\
\sin\left(\frac{\pi}{M}\right) &\;& \cos\left(\frac{\pi}{M}\right)
\end{array}
\right]\quad ,
\end{equation}
and it represents a linear transformation given by a
counterclockwise rotation through angle $\pi/M$. We assume that the
reference density operator has the general form
\begin{equation}
\rho_0=\left[
\begin{array}{ccc}
\alpha &\;& \beta\\
\beta &\;& 1-\alpha
\end{array}
\right]\;,
\end{equation}
where $\alpha$ and $\beta$ are real numbers. Since $\rho_0$ is PSD,
the feasible values of $\alpha$ and $\beta$ are constrained as
$0\leq \alpha \leq 1$ and $|\beta|\leq \sqrt{\alpha (1-\alpha)}$,
respectively. Without loss of generality we assume $\alpha\geq 1/2$.

The operator $S$ has two non-degenerate eigenvalues equal to
$\lambda_1(S)=e^{i\,\pi/M}$ and $\lambda_2(S)=e^{-i\,\pi/M}$.
Therefore, the corresponding eigenvectors define the basis
\begin{equation}
U=\frac{1}{\sqrt{2}}\left[
\begin{array}{ccc}
1 &\;& 1\\
-i &\;& i
\end{array}
\right]\;.
\end{equation}
Consequently,  for the dual problem \texttt{DP3} the optimization
operator $\tilde{X}$ has to be diagonal
\begin{equation}
\tilde{X}=\left[
\begin{array}{ccc}
\tilde{x}_1 &\;& 0\\
0 &\;& \tilde{x}_2
\end{array}
\right]\;,
\end{equation}
where, both $\tilde{x}_1$ and $\tilde{x}_2$ are real and
non-negative, being $\tilde{X}$ PSD. The constraint $\tilde{X} \geq
\frac{1}{M} U^\dag \rho_0 U$ reads
\begin{equation}
\tilde{X} \geq \frac{1}{2M}\left[
\begin{array}{ccc}
1 &\;& (2\alpha-1)+i \, 2\beta\\
 (2\alpha-1)-i \, 2\beta &\;& 1
\end{array}
\right]\;,
\end{equation}
and after some simple algebra,\footnote{It is recalled that an
Hermitian matrix is PSD if and only if its principal minors are all
non-negative.} we find that it can be rewritten as
\begin{equation}
(2M\tilde{x}_1-1)(2M\tilde{x}_2-1)-[(2\alpha-1)^2+(2\beta)^2]\geq
0\quad,
\end{equation}
with $\tilde{x}_1\geq {1}/({2M})$ and $\tilde{x}_2\geq
{1}/({2M})\,$. Hence, the minimum of $\Tr(\tilde{X})=\tilde{x}_1 +
\tilde{x}_2$ is obtained for
$\tilde{x}_1=\tilde{x}_2=1/(2M)(1+\sqrt{(2\alpha-1)^2+(2\beta)^2}$.
In conclusion, the minimum error probability $P_e=1-P_c= 1-
\Tr(\tilde{X})$ is
\begin{equation}
P_e=\frac{M-1}{M} - \frac{1}{M}
\sqrt{(2\alpha-1)^2+(2\beta)^2}\espace. \label{eq:15}
\end{equation}
We note that the first term on the right-hand side of \eqref{eq:15}
corresponds to a blind guessing on the equiprobable elements
belonging to the quantum state ensemble, while the second term
represents the gain due to the optimal quantum discrimination. It is
interesting to observe that the minimum error probability is
obtained for  $\beta=\sqrt{\alpha (1-\alpha)}$ and it results
\begin{equation}
P_e= 1- \frac{2}{M}\espace. \label{eq:16}
\end{equation}
For instance, \eqref{eq:16} holds for $\alpha=1$ and $\beta=0$ which
is the case of study considered by Helstrom \cite{Helstrom} and Ban
\textit{et al.} \cite{Ban}, which models linearly dependent
spin--1/2 quantum states, where the generating density operator has
rank-one $\rho_0= \k(\psi_0)\b(\psi_0)$ with pure state
${\k(\psi_0)=\scriptsize \Big[
\begin{array}{c}
1\\
0
\end{array} \Big]}\,$.
When $\rho_0$ is diagonal, i.e., $\beta=0$, \eqref{eq:15} simplifies
as
\begin{equation}
P_e=1 -\alpha  \frac{2}{M}\,\espace . \label{eq:17}
\end{equation}

The optimal reference measure operator $\Pi_0$ can be found from the
conditions given in Section \ref{Sect:3c} $(U
\tilde{X}_{\textrm{opt}} U^\dag-\frac{1}{M}\rho_0)\Pi_0=\Pi_0 (U
\tilde{X}_{\textrm{opt}} U^\dag-\frac{1}{M}\rho_0) = 0$, that in
this example simplify as
$(\tilde{x}_1\,I-\frac{1}{M}\rho_0)\Pi_0=\Pi_0 (
\tilde{x}_1\,I-\frac{1}{M}\rho_0) = 0$, being
$\tilde{X}_{\textrm{opt}}=\tilde{x}_1 \,I$. Note that these
conditions also imply that $\rho_0 \Pi_0=\Pi_0 \rho_0\,$. The
optimal $\Pi_0$ can then be numerically found by solving a linear
system of equations including the additional requirements $\Pi_0\in
\mathcal{P}$ and $\sum_{i=0}^{M-1} S^i \, \Pi_0 \,S^{-i}= {I}\,$.

It is useful to observe that the condition ${\sum_{i=0}^{M-1} S^i \,
\Pi_0 \,S^{-i}= {I}}$ implicitly fixes the value of the trace of
$\Pi_0$. In fact, $\Tr(\sum_{i=0}^{M-1} S^i \, \Pi_0 \,S^{-i})=
\sum_{i=0}^{M-1}\Tr( S^i \, \Pi_0 \,S^{-i}) = \sum_{i=0}^{M-1}\Tr(
\Pi_0 \,S^{-i}S^i )= M \Tr(\Pi_0)$ and being ${\Tr(I)=N}$ it follows
that $\Tr(\Pi_0)=N/M\,$. By simple algebra it can be found a
closed-form expression for $\Pi_0$ in the following two cases.
\begin{description}
\item[\normalfont 1.] $\beta=0\,$. The optimal $\Pi_0$ is given by
\begin{equation}
\Pi_0= \frac{2}{M}\left[
\begin{array}{cc}
1 & 0 \\
0 & 0
\end{array}
\right]\espace.
%
%
\end{equation}
In particular, setting $\alpha=1/3$ the same numerical results
obtained in \cite{Chou} are found.

\item[\normalfont 2.] $\beta=\sqrt{\alpha (1-\alpha)}\,$.
The measure operator results
\begin{equation}\label{eq:20}
\Pi_0= \frac{2}{M}\, \rho_0 \quad .
\end{equation}
The constraint ${\sum_{i=0}^{M-1} S^i \, \Pi_0 \,S^{-i}= {I}}$
becomes $\frac{2}{M}\sum_{i=0}^{M-1} \rho_i = {I}$ showing that the
quantum state ensemble has a particular structure. Indeed, such a
specific geometry has been considered by Yuen \textit{et al.} in
\cite[(IV.4)]{Yuen} and the results therein reported are in
agreement with \eqref{eq:16}  and \eqref{eq:20}$\,$.
\end{description}


The proposed formulation of the dual problem has also proved useful
to numerically solve systems of large dimensions, where the
computational complexity sets a severe limit to the possibility of
finding an optimal solution. This is the case of pulse position
modulation (PPM), a modulation format candidate for deep space
communications \cite{JPL}. In quantum PPM the states are defined in
a Hilbert space given by the tensorial product of $M$ subspaces,
each of dimension $n$, and, therefore, the overall space dimension
$N$ grows exponentially with the PPM order $M$, being $N={n}^M$
\cite{Cariolaro1}. In \cite{Cariolaro2}, \texttt{DP3} is applied to
quantum PPM using the software \texttt{CVX} for SDP \cite{Grant}.
The solution of \texttt{DP3} results considerably faster than
\texttt{DP1}. As a limit case, for values of $N$ about 1000 the
discrimination problem was successfully solved with \texttt{DP3}
while, on the same processing unit, \texttt{CVX} fails with
\texttt{DP1}, because of computer memory limits.


\section{Conclusions}\label{Sect:5}

We have studied the dual problem for minimum error probability
discrimination of symmetric quantum states. It has been shown that
the optimization operator, in the objective function, can be assumed
to commute with the symmetry operator. This result leads to an
alternative formulation of the dual problem, that presents a reduced
number of variables and constraints. The obtained dual statement is
convenient to find analytical solutions to the discrimination
problem, as we showed with an illustrative example. On the other
hand, the new formulation also permits a computationally efficient
numerical solution by means of SDP methods. This property is
particularly useful to study the performance limits of quantum
mechanical systems described by geometrically uniform quantum states
on Hilbert spaces of large dimensions, such as modulated coherent
states in optical communications.

\begin{acknowledgments}
The authors would like to thank M. Sasaki for encouraging comments.
This work was supported in part by the Q-FUTURE project (prot.
STPD08ZXSJ), University of Padua.
\end{acknowledgments}


\end{document}